\documentclass[10pt,conference]{IEEEtran}\IEEEoverridecommandlockouts
\usepackage{algorithmic}
\usepackage{url}

\usepackage{optidef} % for optimization problem
\usepackage{cite}
\usepackage{amsmath,amssymb,amsfonts}
\usepackage{bm}
\usepackage{algorithm}
\usepackage{algorithmic}

\usepackage{float}
\usepackage{amsthm}
\usepackage{graphics}
\usepackage{graphicx}
\usepackage{textcomp}
\usepackage{xcolor}
\usepackage{dsfont}
\usepackage{comment}
\usepackage{rotating}
\usepackage{makecell}
\usepackage{multirow}

\theoremstyle{plain}

 \usepackage[moderate,tracking=normal]{savetrees}
\def\BibTeX{{\rm B\kern-.05em{\sc i\kern-.025em b}\kern-.08em
    T\kern-.1667em\lower.7ex\hbox{E}\kern-.125emX}}
\usepackage{subcaption}    
	\definecolor{mygreen}{rgb}{0.01, 0.75, 0.24}    %\IEEEoverridecommandlockouts\IEEEpubid{\makebox[\columnwidth]{ 978-1-6654-3540-6/22~\copyright~2022 IEEE \hfill} \hspace{\columnsep}\makebox[\columnwidth]{ }}

\begin{document}

\title{Detecting Multiple Targets with Distributed Sensing and Communication in Cell-Free Massive MIMO

\author{\IEEEauthorblockN{
Zinat Behdad\IEEEauthorrefmark{1}, \"Ozlem Tu\u{g}fe Demir\IEEEauthorrefmark{2}, Ki Won Sung\IEEEauthorrefmark{1}, and Cicek Cavdar \IEEEauthorrefmark{1}
}

\IEEEauthorblockA{\IEEEauthorrefmark{1}Department of Computer Science, KTH Royal Institute of Technology, Stockholm, Sweden\\
(\{zinatb, sungkw, cavdar\}@kth.se)
}

\IEEEauthorblockA{\IEEEauthorrefmark{2}Department of Electrical-Electronics Engineering, TOBB ETU, Ankara, Türkiye  (ozlemtugfedemir@etu.edu.tr)
}
}
\thanks{This work is part of Celtic-Next project RAI-6Green: Robust and AI Native 6G for Green Networks with project-id: C2023/1-9 and 6G-SUSTAIN: Sensing Integrated Elastic 6G Networks for Sustainability. Both projects are funded by Vinnova in Sweden.  \"O. T. Demir was supported by 2232-B International Fellowship for Early Stage Researchers Programme funded by the Scientific and Technological Research Council of T\"urkiye.}
}
\maketitle
\begin{abstract}
This paper investigates multi-target detection in an integrated sensing and communication (ISAC) system within a cell-free massive MIMO (CF-mMIMO) framework. We adopt a user-centric approach for communication user equipments (UEs) and a distributed sensing approach for multi-target detection. A heuristic access point (AP) mode selection algorithm and a channel-aware distributed sensing scheme are proposed, where local measurements at receive APs (RX-APs) are weighted based on the received signal's signal-to-interference ratio (SIR). A maximum a posteriori ratio test (MAPRT) detector is applied under two awareness levels at RX-APs. To balance the communication-sensing trade-off, we develop a power allocation algorithm to jointly maximize the minimum detection probability and communication signal-to-interference-plus-noise ratio (SINR) while satisfying power constraints. The proposed scheme outperforms non-weighted methods. Adding test statistics from more RX-APs can degrade sensing performance due to weaker channels, but this effect can be mitigated by optimizing the weighting exponent. Additionally, assigning more sensing RX-APs to a sensing area results in approximately 10\,dB loss in minimum communication SINR due to limited communication resources.
\end{abstract}
%\vspace{-2mm}
\section{Introduction}
Integrated sensing and communication (ISAC) and cell-free massive MIMO (CF-mMIMO) are key technologies for 6G networks \cite{liu2022integrated}. CF-mMIMO, with its distributed access points (APs), supports advanced sensing networks through multi-static sensing, avoiding the full-duplex requirement of mono-static systems. The cloud radio access network (C-RAN) architecture enables coordinated processing across APs, facilitating ISAC functionalities on general-purpose processors \cite{demir2024cell}. Traditional CF-mMIMO systems, where all APs serve all user equipments (UEs), face scalability challenges due to high computational demands. This issue worsens in CF-mMIMO ISAC systems, where centralized sensing and multi-target detection strain radio and processing resources. To address scalability, \cite{8000355} and \cite{cell-free-book} consider a user-centric communication approach, assigning UEs to a limited number of APs. Building on this, \cite{buzzi2024scalability} introduces a target-centric sensing approach, where APs are allocated to sensing areas based on target locations, enhancing scalability and efficiency.

Multi-target detection in CF-mMIMO requires distinguishing reflections from multiple targets. While joint detection via multiple hypothesis testing is accurate \cite{zhang2023multi-target}, it is computationally prohibitive for large systems. A practical alternative is distributed sensing, where APs locally process signals before forwarding them to the cloud, reducing computational load and enabling real-time detection.

%Moreover, the 3GPP service requirements for ISAC suggest energy-efficient sensing by disabling inactive transmitters and receivers \cite{3gpp_ts_22_137}. Integrating these energy-efficient strategies with distributed sensing could enhance the sustainability of large-scale CF-mMIMO deployments.
%These considerations highlight the need for resource-efficient architectures, motivating this paper to explore practical solutions for CF-mMIMO ISAC systems in next-generation wireless networks.

Distributed sensing for multi-target detection is studied in \cite{elfiatoure2024multiple}, focusing on AP mode selection and power control to optimize communication spectral efficiency (SE) while ensuring a specific mainlobe-to-average-sidelobe ratio (MASR) for sensing zones. Unlike previous works where APs serve either communication or sensing, we consider an integrated approach where transmit APs (TX-APs) handle both downlink communication and sensing, while receive APs (RX-APs) capture target reflections. Existing studies overlook the challenge of aggregating distributed measurements at the cloud, essential for real-time ISAC systems. Building on \cite{Zou2024Distributed}, which addressed single-target detection, we focus on multi-target scenarios requiring higher resources to ensure high detection probability and low false alarm probability while balancing communication and sensing performance.

%Distributed sensing for multi-target detection is studied in \cite{elfiatoure2024multiple}, focusing on AP mode selection schemes and power control strategies to optimize communication spectral efficiency (SE) while ensuring a specific mainlobe-to-average-sidelobe ratio (MASR) for sensing zones. In that work, APs operate either as communication APs or sensing APs. In contrast, we consider a more integrated approach, where transmit APs (TX-APs) handle both downlink communication and sensing transmissions, while receive APs (RX-APs) are dedicated to capturing target reflections. 
%Existing works do not address the challenge of aggregating distributed measurements at the cloud, which is crucial for real-time ISAC systems. To tackle this, we focus on distributed sensing for multi-target detection in CF-mMIMO systems. In \cite{Zou2024Distributed}, we introduced a framework for single-target scenarios, prioritizing detection probability and minimizing fronthaul signaling. However, multi-target detection remains challenging, requiring higher resources to provide high detection probability and low false alarm probability for all targets while balancing the trade-off between communication and sensing performance.

This paper investigates distributed sensing for multi-target detection in a user-centric CF-mMIMO system, where each sensing area is served by limited TX-APs and RX-APs. The key research question is: (RQ) How should local measurements from RX-APs be aggregated at the central cloud? The accuracy of local test statistics at RX-APs depends on sensing signal-to-interference-plus-noise ratio (SINR)—lower SINR reduces detection accuracy, while higher SINR degrades communication performance. We aim to balance both by maximizing the minimum sensing SINR ($\gamma_s$) and communication SINR ($\gamma_c$). The main contributions of this paper are:
\begin{itemize} 
    \item We propose a heuristic AP mode selection scheme and a distributed sensing approach that weights local test statistics based on signal-to-interference ratio (SIR) and channel quality. We derive test statistics using a maximum a posteriori ratio test (MAPRT) detector under two RX-AP awareness levels: i) fully-informed system (FIS) with full transmit signal access and ii) partially-informed system (PIS) with only statistical knowledge.

    \item We develop a power allocation algorithm that jointly maximizes the minimum sensing SINR and communication SINR per UE while meeting power constraints, by solving the non-convex problem via convex-concave programming (CCP).  

    \item We analyze minimum detection probability in multi-target vs. single-target scenarios, assess the impact of target reflectivity (radar cross section (RCS)), and examine how the number of TX-APs, RX-APs, and weighting factors affect sensing and communication performance.  

\end{itemize}

\begin{figure}[tbp]
\centerline{\includegraphics[trim={0mm 0mm 0mm 0mm},clip,
width=0.9\linewidth]{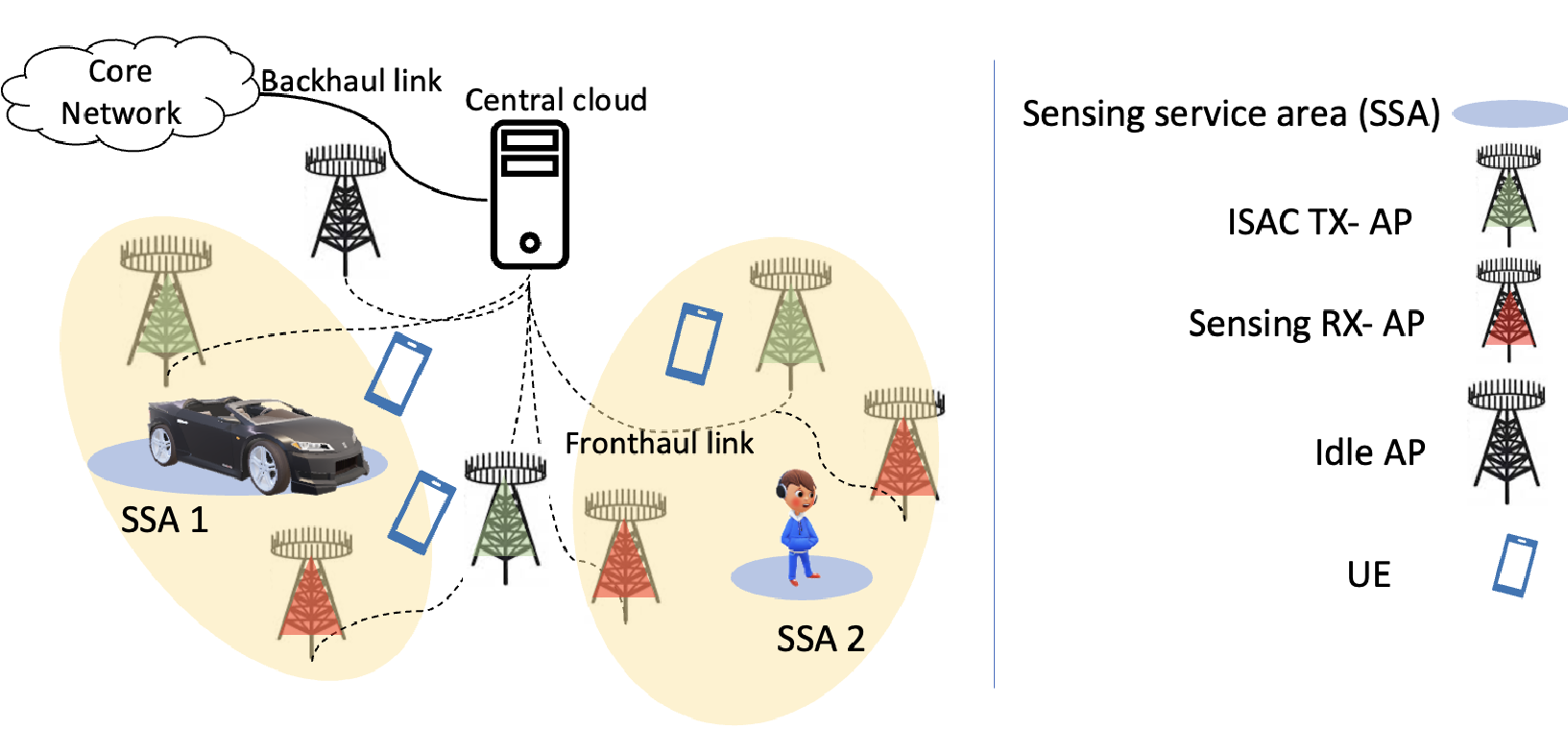}}
 %\vspace{-2mm}
\caption{Multi-target ISAC system model in CF-mMIMO.}
\label{fig1}%\vspace{-5mm}
\end{figure}
%\vspace{-1mm}
\section{System Model}\label{sec:system model}
We consider a CF-mMIMO ISAC system with distributed operation within a C-RAN architecture, as shown in Fig.~\ref{fig1}. The number of APs is $L$, each equipped with $M$ antennas deployed in a horizontal uniform linear array (ULA) with half-wavelength spacing. APs are connected to the cloud via fronthaul links and are either operating in ISAC transmission, sensing reception, or idle mode. There are $K$ single-antenna UEs and $S$ non-overlapping target sensing service areas (SSAs).

We define $\mathcal{L}_{\rm tx}$, $\mathcal{L}_{\rm rx}$, $\mathcal{L}_{\rm idle}$, $\mathcal{K}$, and $\mathcal{S}$, as the set of ISAC TX-APs, sensing RX-APs, idle APs, UEs and target SSAs, respectively. We define $\mathcal{M}_{k}$ as the set of TX-APs assigned to serve UE $k$, and  $\mathcal{T}_{s}\subset \mathcal{L}_{\rm tx}$ and $  {\mathcal{R}}_{s}\subset \mathcal{L}_{\rm rx}$ as the sets of TX-APs and RX-APs assigned to SSA $s$, respectively. Moreover, $\mathcal{U}_{l}$ is the set of UEs and $\mathcal{S}_{l}$ the set of targets which are assigned to TX-AP $l$. The cardinality of the sets  $\mathcal{L}_{\rm tx}$ and $\mathcal{L}_{\rm rx}$ are denoted by $L_{\rm tx}$ and $L_{\rm rx}$, respectively.
 
%\subsection{Downlink Data and Sensing Transmission}
The transmit signal at AP $l$ is given by 
\begin{align}
    \textbf{x}_l[m]&= \sum_{k\in \mathcal{U}_l}  \sqrt{p_{k,l}}\,\textbf{w}_{k,l}\, s_k[m]\,+\sum_{s\in \mathcal{S}_l}\sqrt{q_{s,l}}\,  \boldsymbol{\omega}_{s,l}\, r_s[m],%\vspace{-2mm}%\nonumber\\
   % & = \textbf{W}^{\text{comm}}_l \textbf{S}[m] \boldsymbol{\rho}'_l + \textbf{W}^{\text{sens}}_l \textbf{R}[m] \boldsymbol{q}'_l
\end{align}
where $p_{k,l}$ and $q_{s,l}$ are the power coefficients for UE $k$ and SSA $s$, respectively. The communication and sensing symbols for UE $k$ and SSA $s$ for the time-frequency channel use $m$ are defined as $s_k[m]$ and $r_s[m]$. The precoding vectors $\textbf{w}_{k,l}$ and $\boldsymbol{\omega}_{s,l}$ are obtained based on the channel state information (CSI) with a distributed operation using local partial minimum mean-squared error (LP-MMSE) and maximum ratio transmission (MRT) precoding approaches, respectively \cite{cell-free-book}.

We define $\textbf{h}_{kl}\in \mathbb{C}^M$ as the channel vector from AP $l$ to UE $k$ which is modeled as spatially correlated Rician fading as $\textbf{h}_{kl} =e^{j\psi_{kl}}\bar{\textbf{h}}_{kl}\,+\,\tilde{\textbf{h}}_{kl}$,
consisting of a semi-deterministic line-of-sight (LOS) path, represented by $e^{j\psi_{kl}}\bar{\textbf{h}}_{kl}$ with unknown phase-shift $\psi_{kl}\sim \mathcal{U}[0,2\pi)$, i.e., uniformly distributed on $[0,2\pi)$, and a stochastic non-LOS (NLOS) component $\tilde{\textbf{h}}_{kl}\sim \mathcal{CN}(\textbf{0},\tilde{\textbf{R}}_{kl})$ with the spatial correlation matrix $\tilde{\textbf{R}}_{kl}\in \mathbb{C}^{M\times M}$. Both $\bar{\textbf{h}}_{kl}$ and $\tilde{\textbf{R}}_{kl}$ include the effect of geometric path loss. 

We use LP-MMSE precoding approach for communication UEs as presented in \cite{cell-free-book}. 
The normalized local transmit precoding vector for UE $k$ at the TX-AP $l$, denoted by $\textbf{w}_{k,l}$, is obtained as $\textbf{w}_{k,l}= \frac{\overline{\textbf{w}}_{k,l}}{\sqrt{\mathbb{E}\{\Vert \overline{\textbf{w}}_{k,l}\Vert^2\}}}$, where
$\overline{\textbf{w}}_{k,l}=p_k^{\text{ul}} \left( \sum_{i\in {\mathcal{U}}_l} p_i^{\text{ul}}\left(\hat{\textbf{h}}_{il}\hat{\textbf{h}}_{il}^H + \textbf{Z}_{il}\right)+ \sigma_n^2 \textbf{I}_M\right)^{-1} \hat{\textbf{h}}_{kl}$ and $p_i^{\text{ul}}$ is the pilot transmit power of UE $i$ in uplink channel estimation phase, $\hat{\textbf{h}}_{kl}$ is the linear minimum mean-squared error (LMMSE) channel estimate of the communication channel $\textbf{h}_{kl}$, as obtained in \cite{wang2020uplink}, %$\textbf{Z}_{il}= \textbf{R}_{il}-p_i^{\rm ul}\,\tau_p\, \textbf{R}_{il}\left(\boldsymbol{\Psi}'_{t_i,k}\right)^{-1}\textbf{R}_{il}$
$\textbf{Z}_{il}$ is the correlation matrix of the channel estimation error $\textbf{h}_{il}-\hat{\textbf{h}}_{il}$, and $\sigma_n^2$ is the noise variance.
%The LMMSE channel estimate of $\textbf{h}_{kl}$ is
%\begin{align}\label{eq:hHat}
%\hat{\textbf{h}}_{k,l}= \sqrt{p_k^{\rm ul}}\textbf{R}_{k,l}\left(\boldsymbol{\Psi}_{t_k,l} \right)^{-1}\textbf{y}_{t_k,l}^{p},
%\end{align}

The MRT precoding vectors for sensing are obtained as $\boldsymbol{\omega}_{s,l}=\frac{\boldsymbol{\mathfrak{h}}^*_{sl}}{\left \Vert \boldsymbol{\mathfrak{h}}_{sl}\right \Vert}$, where $\boldsymbol{\mathfrak{h}}_{sl}=\sqrt{\beta_{sl}}\textbf{a}(\varphi_{s,l},\vartheta_{s,l})$ denotes the LOS channel between AP $l$ and the designated SSA \(s\). $\beta_{sl}$ is the channel gain corresponding to the TX-target path and $\textbf{a}(\varphi_{s,l},\vartheta_{s,l}) =\begin{bmatrix}
          1& e^{j\pi \sin(\varphi_{s,l})\cos(\vartheta_{s,l})}& \ldots& e^{j(M-1)\pi\sin(\varphi_{s,l})\cos(\vartheta_{s,l})}
        \end{bmatrix} ^T $ is the array response vector with the azimuth and elevation angles from TX-AP $l$ to the target location $s$, denoted by $\varphi_{s,l}$ and $\vartheta_{s,l}$ respectively. The NLOS paths are disregarded in this model due to the considerable signal attenuation associated with multiple path loss effects in a two-way sensing channel\cite{buzzi2024scalability}.
The received signal at UE $k$ is given as
\begin{align}
    y_{k}[m]\!%&= \sum_{l\in\mathcal{L}_{\rm tx}} \textbf{h}_{kl}^H \textbf{x}_l[m] +n_k[m] \nonumber\\
    & =\!\sum_{l\in\mathcal{M}_{k}} \!\sqrt{p_{k,l}} \textbf{h}_{kl}^H\!\textbf{w}_{k,l}\! s_k[m]\! \nonumber\\
    &\quad+  \sum_{l\in\mathcal{L}_{\rm tx}}\!\sum_{j\in \mathcal{U}_l\setminus \{k\}}\!\sqrt{p_{j,l}}\textbf{h}_{kl}^H\textbf{w}_{j,l}\, s_j[m]\,\nonumber\\
    &\quad +\sum_{l\in\mathcal{L}_{\rm tx}}   \sum_{s\in \mathcal{S}_l} \sqrt{q_{s,l}}\,  \textbf{h}_{kl}^H\,\boldsymbol{\omega}_{s,l}\, r_s[m] + n_k[m],
\end{align}
where $n_k[m]\sim \mathcal{CN}(0,\sigma_n^2)$ is the receive noise at UE $k$. The effective SINR for UE $k$ leading to the downlink achievable SE given in \cite{demir2024cell} is given by
%The downlink achievable SE of UE $k$ is given by \cite{demir2024cell} as\begin{align}\mathsf{SE}_k = \frac{\tau_d}{\tau_c} \log_2 \left( 1+ \mathsf{SINR}^{\text{comm}}_k\right) \quad \text{bit/s/Hz}\end{align}where $\mathsf{SINR}^{\text{comm}}_k$ is the effective signal-to-noise ratio (SINR) and $\tau_d$ and $\tau_c$ are the number of symbols in downlink transmission and one coherence block, respectively. 
\begin{align}
    \mathsf{SINR}^{\text{comm}}_k = \frac{\vert \textbf{a}_k^T \boldsymbol{\rho}_k\vert ^2}{\sum_{j\in\mathcal{K}}\boldsymbol{\rho}_j^T \textbf{B}_{kj}\boldsymbol{\rho}_j+ \sum_{s\in \mathcal{S}}\boldsymbol{q}_s^T \textbf{C}_{ks}\boldsymbol{q}_s + \sigma_n^2},
\end{align}
where $\boldsymbol{\rho}_k=\begin{bmatrix} \sqrt{p_{k,1}}& \ldots & \sqrt{p_{k,L_{\rm tx}}}\end{bmatrix}^T \in \mathbb{R}^{L_{\rm tx}}$, $\boldsymbol{q}_s=\begin{bmatrix} \sqrt{q_{s,1}}& \ldots & \sqrt{q_{s,L_{\rm tx}}}\end{bmatrix}^T\in \mathbb{R}^{L_{\rm tx}} $ are the concatenated power coefficient vectors from all TX-APs corresponding to UE $k$ and SSA $s$, respectively. The entries of $\textbf{a}_k \in \mathbb{C}^{L_{\rm tx}} $, $\textbf{B}_{kj}\in \mathbb{C}^{L_{\rm tx}\times L_{\rm tx}}$, and $\textbf{C}_{ks}\in \mathbb{C}^{L_{\rm tx}\times L_{\rm tx}}$ are 
\begin{align}
    &[\textbf{a}_k]_l = \mathbb{E}\left\{\textbf{h}_{kl}^H \textbf{w}_{k,l}\right\}\geq 0 \\
    &[\textbf{B}_{kj}]_{ll'} = \left\{\begin{matrix}
     \mathbb{E}\left\{\textbf{h}_{kl}^H\, \textbf{w}_{j,l}\, \textbf{w}_{j,l'}^H\,\textbf{h}_{kl'}\right\}& k\neq j \\
    \mathbb{E}\left\{\textbf{h}_{kl}^H\, \textbf{w}_{j,l}\, \textbf{w}_{j,l'}^H\,\textbf{h}_{kl'}\right\}-[\textbf{a}_k]_l\,[\textbf{a}_k]_{l'}^*& k=j   \\
    \end{matrix}\right. \\
    & [\textbf{C}_{ks}]_{ll'} = \mathbb{E}\left\{\textbf{h}_{kl}^H\,\boldsymbol{\omega}_{s,l}\,\boldsymbol{\omega}_{s,l'}^H\,\textbf{h}_{k,l'} \right\}.  
\end{align}
\subsection{Sensing Reception}%\vspace{-1mm}
The sensing RX-APs jointly receive the reflected signals from the targets in the SSAs and obtain the local test statistics corresponding to the assigned SSAs. The local test statistics then will be sent to the cloud to be processed for making decision about the presence of the target at each SSA. Although each sensing RX-AP may receive reflections from all targets, it processes only the signals corresponding to its assigned targets to reduce computational complexity and focus on relevant sensing tasks.\footnote{We assume a known TX-RX channel without targets and neglect communication interference, leaving its impact for future work.}
To maximize the received signal quality, we employ normalized maximum ratio combining (MRC) vectors. Assume RX-AP $r$ is assigned to SSA $s$, i.e., $r\in\mathcal{R}_s$.  The MRC vector at RX-AP $r$ corresponding to SSA $s$ is defined as $\textbf{v}_{s,r}=\frac{\boldsymbol{\mathsf{h}}_{sr}}{\left \Vert \boldsymbol{\mathsf{h}}_{sr}\right\Vert}$, where $\boldsymbol{\mathsf{h}}_{sr}=\sqrt{\overline{\beta}_{sr}}\textbf{a}(\phi_{s,r},\theta_{s,r})\in \mathbb{C}^{ M}$ with channel gain $\overline{\beta}_{sr}$ and azimuth and elevation angle $\phi_{s,r}$ and $\theta_{s,r}$, respectively. After applying the combining vector, the received signal at AP $r$ with respect to SSA $s$, is given as
\begin{align} \label{eq:y_ls}
    y_{s,r} [m]\!&= \!\underbrace{\textbf{v}_{s,r}^H\!\sum_{l\in \mathcal{L}_{\rm tx}}\!\alpha_{s,r,l}\!\textbf{G}_{s,r,l}\textbf{x}_l[m]}_{\triangleq g_{s,r}[m], \text{desired signal}} \nonumber\\
    &+\!\underbrace{\textbf{v}_{s,r}^H\!\sum_{t\in \mathcal{S}\setminus \{s\} }\!\sum_{l\in \mathcal{L}_{\rm tx}}\!\alpha_{t,r,l}\textbf{G}_{t,r,l}\textbf{x}_l[m]}_{\text{target-related interference}}
   + \!\underbrace{\textbf{v}_{s,r}^H\textbf{n}_r[m]}_{\triangleq n_{s,r}[m], \text{noise}}
\end{align}
where $\alpha_{s,r,l}\sim\mathcal{CN}(0,1)$ is the normalized RCS following Swerling-I model, and $\textbf{G}_{s,r,l}=\sqrt{\beta_{s,r,l} }\textbf{a}(\phi_{s,r},\theta_{s,r})\textbf{a}^{T}(\varphi_{s,l},\vartheta_{s,l})\in \mathbb{C}^{M \times M}$ is the two-way channel matrix from TX-AP $l$ to RX-AP $r$ through the SSA $s$, where $\beta_{s,r,l}$ is the channel gain of the path between TX-AP $l$ and RX-AP $r$ through SSA $s$. Channel gain contains path loss and RCS variance of the target. 
Let us define matrix $\textbf{G}_{s,r}[m]=\begin{bmatrix}\textbf{G}_{s,r,1}\textbf{x}_1[m]&\cdots&\textbf{G}_{s,r,L_{\rm tx}}\textbf{x}_{L_{\rm tx}}[m]\end{bmatrix}\in \mathbb{C}^{M\times L_{\rm tx}}$, and vector $\boldsymbol{\alpha}_{s,r}\in \mathbb{C}^{L_{\rm tx}}$ containing the $\alpha_{s,r,l}$.
We can write the equation~\eqref{eq:y_ls} in a more compact matrix form as
\begin{align}
     y_{s,r}[m]\! &=\!\textbf{v}_{s,r}^H\,\textbf{G}_{s,r}[m]\boldsymbol{\alpha}_{s,r}
     \!+\!\textbf{v}_{s,r}^H\!\!\!\sum_{t\in\mathcal{S}\setminus \{s\}}\!\!\textbf{G}_{t,r}[m]\boldsymbol{\alpha}_{t,r}
    \!+ \!n_{s,r}[m].
\end{align}

The sensing SINR at RX-AP $r$ and for SSA $s$ over $\tau_s$ channel uses is obtained as in \eqref{eq:sensing_snr}, on the top of next page,
\begin{figure*}
\begin{align}
\label{eq:sensing_snr}
    \mathsf{SINR}^{\text{sens}}_{s,r} = \frac{ \sum_{m=1}^{\tau_s}\sum_{l\in \mathcal{L}_{\rm tx}}\vert\textbf{d}_{s,r,l,m}^T\overline{\boldsymbol{\rho}}_l+\textbf{e}_{s,r,l,m}^T\overline{\boldsymbol{q}}_l\vert^2}{\sum_{m=1}^{\tau_s}\sum_{t\in \mathcal{S}\setminus \{s\}}\sum_{l\in \mathcal{L}_{\rm tx}}\vert\textbf{f}_{s,t,r,l,m}^T\overline{\boldsymbol{\rho}}_l+\textbf{g}_{s,t,r,l,m}^T\overline{\boldsymbol{q}}_l\vert^2+
    \tau_s\sigma_n^2}.
\end{align} 
   \hrulefill
   %\vspace{-6mm}
\end{figure*}
where $\overline{\boldsymbol{\rho}}_l=\begin{bmatrix} \sqrt{p_{1,l}}& \ldots & \sqrt{p_{K,l}}\end{bmatrix}^T \in \mathbb{R}^{K}$ and $\overline{\boldsymbol{q}}_l=\begin{bmatrix} \sqrt{q_{1,l}}& \ldots & \sqrt{q_{S,l}}\end{bmatrix}^T\in \mathbb{R}^{S}$ are the vectors of communication and sensing power coefficients at TX-AP $l$, respectively and the entries of the vectors $\textbf{d}_{s,r,l,m}$, $\textbf{e}_{s,r,l,m}$, $\textbf{f}_{s,t,r,l,m}$, and $\textbf{g}_{s,t,r,l,m}$ are derived as 
\begin{align}
    &[\textbf{d}_{s,r,l,m}]_{k} = \textbf{v}_{s,r}^H\,\textbf{G}_{s,r,l}\textbf{w}_{k,l}s_k[m], \quad\forall k\in \mathcal{K}\\
    &[\textbf{e}_{s,r,l,m}]_{t} = \textbf{v}_{s,r}^H\,\textbf{G}_{s,r,l}\boldsymbol{\omega}_{t,l}r_{t}[m], \quad \forall t\in \mathcal{S}\\
    &[\textbf{f}_{s,t,r,l,m}]_{k}=\textbf{v}_{s,r}^H\,\textbf{G}_{t,r,l}\textbf{w}_{k,l}s_k[m], \quad\forall k\in \mathcal{K}\\ 
  &[\textbf{g}_{s,t,r,l,m}]_{t'}=\textbf{v}_{s,r}^H\,\textbf{G}_{t,r,l}\boldsymbol{\omega}_{t',l}r_{t'}[m], \quad \forall t'\in \mathcal{S}
.\end{align}
We define the concatenated vector $\boldsymbol{\rho}=\begin{bmatrix} \overline{\boldsymbol{\rho}}_1^T& \overline{\boldsymbol{q}}_1 ^T& \ldots & \overline{\boldsymbol{\rho}}_{L_{\rm tx}}^T  &\overline{\boldsymbol{q}}_{L_{\rm tx}} ^T\end{bmatrix}^T\in \mathbb{R}^{(K+S)L_{\rm tx}}$, and block diagonal matrices $\boldsymbol{\mathsf{A}}_{s,r}$ and  $\boldsymbol{\mathsf{B}}_{s,r}$ with $l$th blocks
\begin{align}
    &\left[\boldsymbol{\mathsf{A}}_{s,r}\right]_{l}= \sum_{m=1}^{\tau_s} \Re\left\{ \begin{bmatrix}
        \textbf{d}_{s,r,l,m}\\\textbf{e}_{s,r,l,m}
    \end{bmatrix}\begin{bmatrix}
        \textbf{d}_{s,r,l,m}^H&\textbf{e}_{s,r,l,m}^H
    \end{bmatrix}\right\}\\
&[\boldsymbol{\mathsf{B}}_{s,r}]_{l}\!\!\!\!=\!\! \!\!\sum_{m=1}^{\tau_s}\!\!\!\!\!\!\sum_{t\in \mathcal{S}\!\setminus \! \{s\}}\!\!\!\!\!\!\!\!\!\!\Re\!\!\left\{\! \!\begin{bmatrix}
        \textbf{f}_{s,t,r,l,m}\\\textbf{g}_{s,t,r,l,m}
    \end{bmatrix}\left[
        \textbf{f}_{s,t,r,l,m}^H \ \textbf{g}_{s,t,r,l,m}^H\right]
    \!\!\right\}.
\end{align}
Finally, we can rewrite the sensing SINR expression as follows
%\begin{align}
 %   \mathsf{SNR}^{\text{sens}}_{r,s} = \frac{\overline{\boldsymbol{\rho}}^T \textbf{B}\overline{\boldsymbol{\rho}}}{\tau_s\sigma_n^2}
%\end{align}
\begin{align}
    \mathsf{SINR}^{\text{sens}}_{s,r} = \frac{{\boldsymbol{\rho}}^T \boldsymbol{\mathsf{A}}_{s,r}{\boldsymbol{\rho}}}{{\boldsymbol{\rho}}^T \boldsymbol{\mathsf{B}}_{s,r}{\boldsymbol{\rho}}+\tau_s\sigma_n^2}.
\end{align}
\section{Heuristic AP Operation Mode Selection}
We define $S$ and $R$ as the number of TX-APs and RX-APs per SSA and $\mathcal{L}$ as the set of all AP candidates. For each SSA, we first rank APs based on their channel gains and select the AP with the highest gain as the first RX-AP. After determining the first RX-APs across all SSAs, we then select the second-highest APs from the sorted lists as the first TX-APs, ensuring that these APs have not already been assigned as RX-APs. This approach guarantees a TX-RX pair with the highest two-way channel gain, optimizing sensing performance.

Once the first TX- and RX-APs are determined, the remaining 
$R-1$ RX-APs per SSA are selected from the third to the 
$(R+1)$th highest APs in the sorted lists. The total number of RX-APs is bounded by $L_{\rm rx}\leq RS$, as certain APs may be assigned to multiple SSAs. The additional $(T-1)$ TX-APs are chosen from the remaining unassigned APs. For communication UEs, we employ a user-centric approach in which each single-antenna UE is jointly served by a subset of ISAC TX-APs \cite{cell-free-book} where the total channel gain for each UE is above a threshold, denoted by $\beta_\textrm{th}$. The $\beta_{l,k}$ denotes the channel gain from AP $l$ to UE $k$. For each UE, we select the AP with the highest channel gain as the master AP. If an ISAC TX-AP is assigned for both UE communication and target sensing, it simultaneously transmits downlink data and sensing signals toward the target location. As a result, APs in transmission mode may also participate in uplink channel estimation, downlink communication, and/or sensing transmission. The proposed heuristic algorithm is summarized in Algorithm~\ref{algo:AP_selection}.
\begin{algorithm}[t]
    \caption{\textbf{Heuristic AP Mode Selection}}
    \label{algo:AP_selection}
    \begin{algorithmic}[1]
        \STATE \textbf{Input:} Number of TX and RX-APs per SSA ($T,R$), $\mathcal{L}$
        \STATE \textbf{Initialize:} Set $\mathcal{L}_{\rm idle} = \mathcal{L}$ and $\mathcal{L}_{\rm tx}=\mathcal{L}_{\rm rx} =\mathcal{M}_{k}= \mathcal{T}_{s}= \mathcal{R}_{s}= \mathcal{U}_{l}=\mathcal{S}_{l}= \emptyset$
        \FOR{each SSA $s \in \mathcal{S}$}
            \STATE Sort APs in $\mathcal{L}_{\rm idle}$ and assign the AP with the highest gain as the first RX-AP.
            \STATE $\mathcal{R}_{s} \gets \mathcal{R}_{s} \cup \{\text{selected AP}\}$, $\mathcal{L}_{\rm rx} \gets \mathcal{L}_{\rm rx} \cup \{\text{selected AP}\}$.
        \ENDFOR
        \STATE Update $\mathcal{L}_{\rm idle}$: $\mathcal{L}_{\rm idle} \gets \mathcal{L}_{\rm idle} \setminus  \mathcal{L}_{\rm rx}$.
        \FOR{each SSA $s \in \mathcal{S}$}
            \STATE Assign the next highest AP as TX-AP.
            \STATE $\mathcal{T}_{s} \gets \mathcal{T}_{s} \cup \{\text{selected AP}\}$, $\mathcal{L}_{\rm tx} \gets \mathcal{L}_{\rm tx} \cup \{\text{selected AP}\}$.
            \STATE $\mathcal{L}_{\rm idle} \gets \mathcal{L}_{\rm idle} \setminus \{\text{selected AP}\}$.
        \ENDFOR
        \FOR{each SSA $s \in \mathcal{S}$}
            \STATE Select the next $(R-1)$ APs from the sorted list in $\mathcal{L}_{\rm idle}$ (ranked 3 to $R+1$).
            \STATE $\mathcal{R}_{s} \gets \mathcal{R}_{s} \cup \{\text{selected APs}\}$, $\mathcal{L}_{\rm rx} \gets \mathcal{L}_{\rm rx} \cup \{\text{selected APs}\}$.
        \ENDFOR
        \STATE Update $\mathcal{L}_{\rm idle}$: $\mathcal{L}_{\rm idle} \gets \mathcal{L}_{\rm idle} \setminus \mathcal{L}_{\rm rx}$.
        \FOR{each SSA $s \in \mathcal{S}$}
            \STATE Select the remaining $(T-1)$ TX-APs from $\mathcal{L}_{\rm idle} \cup \mathcal{L}_{\rm tx}$.
            \STATE $\mathcal{T}_{s} \gets \mathcal{T}_{s} \cup \{\text{selected APs}\}$, $\mathcal{L}_{\rm tx} \gets \mathcal{L}_{\rm tx} \cup \{\text{selected APs}\}$, $\mathcal{L}_{\rm idle} \gets \mathcal{L}_{\rm idle} \setminus \{\text{selected APs}\}$.
        \ENDFOR
        \FOR{each UE $k \in \mathcal{K}$}
        \STATE Select the AP with highest channel gain $\beta_{l,k}$ as the master AP and update $\mathcal{M}_{k} \gets \mathcal{M}_{k}  \cup\{\text{selected AP} \}$
            \FOR{each AP $l \in \mathcal{L}_{\rm tx} \cup \mathcal{L}_{\rm idle} $}
                \IF{$\sum_{l'\in \mathcal{M}_{k}} \beta_{l',k} < \beta_{\rm th}$}
                    \STATE assign AP $l$ to serve UE $k$: $\mathcal{M}_{k} \gets \mathcal{M}_{k}  \cup\{\text{AP $l$} \}$.
                \ENDIF
            \ENDFOR
        \ENDFOR
        \STATE Update $\mathcal{L}_{\rm tx}$ %\gets \mathcal{L}_{\rm tx} \cup \{\text{selected TX-APs}\}$ 
        and $\mathcal{L}_{\rm idle}$.%\gets \mathcal{L}_{\rm idle} \setminus \{\text{selected TX-APs}\}$.
        \FOR{each TX-AP $l \in \mathcal{L}_{\rm tx}$}
            \STATE Update $\mathcal{U}_{l}$ and $\mathcal{S}_{l}$ with UEs and targerts assigned to TX-AP $l$, respectively.
        \ENDFOR
        \STATE \textbf{return} $\mathcal{L}_{\rm rx}, \mathcal{L}_{\rm tx}, \mathcal{L}_{\rm idle}, \mathcal{M}_{k}, \mathcal{T}_{s}, \mathcal{R}_{s}, \mathcal{U}_{l}, \mathcal{S}_{l}$.
\end{algorithmic}
\end{algorithm}
%\end{comment}

\section{Channel-aware Distributed Detector}
\label{sec:detector}
To detect the multiple targets, we use distributed sensing and form a set of binary hypothesis testing problems for each SSA. The binary hypothesis testing problem for SSA $s$ at RX-AP $r\in \mathcal{R}_s$ is given by 
\begin{align}\label{hypothesis}
   &\mathcal{H}_{s,r,0} : y_{s,r}[m]= I_{s,r}[m]+n_{s,r}[m] \nonumber\\
  &\mathcal{H}_{s,r,1} :y_{s,r}[m]=g_{s,r}[m]+I_{s,r}[m]+n_{s,r}[m].
\end{align}
for $m=1,\ldots, \tau_s$, where $g_{s,r}[m]$,  $I_{s,r}[m]$, and  $n_{s,r}[m]$ are the desired signal, interference and noise signal with respect to the received signal for SSA $s$ and for channel use $m$. $\mathcal{H}_{s,r,0}$ represents the null hypothesis of the target's absence and $\mathcal{H}_{s,r,1}$ represents the alternative hypothesis that the target exists in SSA $s$.

%Each RX-AP obtains local test statistics corresponding to the serving SSAs.
Given $r\in \mathcal{R}_s$, the local test statistic for SSA $s$ at the RX-AP $r$ is denoted by $T_{s,r}$. All local test statistics are aggregated in the cloud to obtain the final test statistic, denoted by $T_s$, which will be used for making a decision about the presence of the target through comparing the final test statistic with a threshold, $\lambda_d$. The threshold $\lambda_d$ is usually selected empirically for a given false alarm probability threshold. 

Let $w_{s,r}$ denote the weight of the local test statistic $T_{s,r}$ which is selected based on the SIR at the RX-AP $r$ for SSA $s$ and the channel quality at this RX-AP compared to the other serving RX-APs, as
\begin{align}\label{eq:weights}
     w_{s,r} = \frac{\overline{w}_{s,r}^v }{\sum_{r'\in \mathcal{R}_s}\overline{w}_{s,r'}^v},
\end{align}
where
$\overline{w}_{s,r} = \frac{\overline{\beta}_{sr}\vert \textbf{v}_{s,r}^H \textbf{a}(\phi_{s,r},\theta_{s,r})\vert^2 }{\sum_{t\in \mathcal{S}\setminus s}\overline{\beta}_{tr}\vert \textbf{v}_{s,t}^H \textbf{a}(\phi_{t,r},\theta_{t,r})\vert^2}$
and weighting exponent $v$ dictates the sensing performance behavior. The final test statistic corresponding to SSA $s$ is
\begin{align}
    T_{s} = \sum_{r\in \mathcal{R}_{s}} w_{s,r} T_{s,r}\begin{matrix}
    \mathcal{H}_{s,1}\\\geq\\<\\ \mathcal{H}_{s,0}
    \end{matrix} \lambda_d.
\end{align}
 
We derive the local test statistics according to the distributed MAPRT detector under two scenarios: i) fully-informed system (FIS), and ii) partially-informed system (PIS).

%\subsection{Correlation-Aware Detector with Fully-Informed System (FIS)}
%\subsection{Distributed Sensing in Fully-Informed System (FIS)}
Let $\textbf{R}_{s,r}^\textrm{rcs}\in \mathbb{C}^{L_\textrm{tx}\times L_\textrm{tx}}$ denote the RCS correlation matrix corresponding to SSA $s$ in the RX-AP $r$. In FIS scenarios, we assume the statistics of the unknown RCS variables (i.e., $\textbf{R}_{s,r}^\textrm{rcs}$) are available and the RX-APs have full access to the transmitted signals. Thus, the local test statistics are\footnote{The local test statistics are suboptimal due to unaccounted sensing interference, which will be addressed in future work. This paper focuses on the aggregation scheme at the cloud.}

\begin{align}
    T^{\rm FIS}_{s,r}= \ln\left(\frac{\max_{\boldsymbol{\alpha}_{s,r}} \prod_{m=1}^{\tau_s}p\left(y_{s,r} [m] |\boldsymbol{\alpha}_{s,r},\mathcal{H}_{s,r,1}\right) p\left(\boldsymbol{\alpha}_{s,r}\right)}{ \prod_{m=1}^{\tau_s}p\left(y_{s,r} [m]|\mathcal{H}_{s,r,0}\right)}\right)
\end{align}
where $p\left(y_{s,r} [m] |\boldsymbol{\alpha}_{s,r},\mathcal{H}_{s,r,1}\right)$ is the probability distribution function (PDF) of the received signal for a given $\boldsymbol{\alpha}_{s,r}$ under hypothesis $\mathcal{H}_{s,r,1}$ and $p\left(\boldsymbol{\alpha}_{s,r}\right)$ is the PDF of RCS variables.
 %The vectorized target-free channel $\boldsymbol{\mathfrak{h}}_{l}$ follows Rayleigh fading distribution with correlation matrix $\normalfont\textbf{R}_l= \normalfont\mathrm{blkdiag}\left(\overline{\textbf{R}}_{l,1},\cdots,\overline{\textbf{R}}_{l,L} \right)$
The test statistic for SSA $s$ at RX-AP $r$ is given by %\eqref{eq:likelihood-MAPRT} on the top of next page.
\begin{align}\label{eq:T_r_FI}
    &T^{\rm MAPRT}_{s,r} = \textbf{a}_{s,r}^H\,\textbf{C}_{s,r}^{-1}\,\textbf{a}_{s,r},\\
    &\textbf{a}_{s,r}= \sum_{m=1}^{\tau_s} \textbf{G}_{s,r}^H[m]\textbf{v}_{s,r} \,y_{s,r}[m]\in \mathbb{C}^{L_{\rm tx}},\\
    &\normalfont\textbf{C}_{s,r}\!= \!\sum_{m=1}^{\tau_s} \!\textbf{G}_{s,r}^H[m]\textbf{v}_{s,r} \textbf{v}_{s,r}^H \textbf{G}_{s,r}[m]\!+ \!\sigma_n^2 (\textbf{R}_{s,r}^\textrm{rcs})^{-1}\in \mathbb{C}^{L_{\rm tx}\times L_{\rm tx}}.
\end{align}
%We call this \emph{correlation-aware} detector since it exploits the off-diagonal entries of $\textbf{C}_{r,s}$ matrix.
%{\subsection{Correlation-Aware Detector with Partially-Informed System (PIS)}
In PIS scenarios, the RCS statistics are available, however the RX-APs have only access to the statistics of the transmit signals. 
Let $c_{s,r,l}[m] \triangleq \,\sqrt{\beta_{s,r,l}}\,\textbf{a}^{T}(\varphi_{s,l},\vartheta_{s,l})\textbf{x}_l[m]$ denote the unknown term at the RX-AP $r$ and define the concatenated vector
$\textbf{c}_{s,r}[m] = \begin{bmatrix} c_{s,r,1}[m]&  \ldots&c_{s,r,L_{\rm tx}}[m]
\end{bmatrix}^H\in \mathbb{C}^{L_{\rm tx}}$.  
The RX-APs need to estimate both RCS values and $\textbf{c}_{s,r}[m]$ as in \eqref{eq:T_PIS}, on the top of next page.
\begin{figure*}
    
\begin{align}\label{eq:T_PIS}
   & T^{\rm PIS}_{s,r}= \ln\left(\frac{\max_{\boldsymbol{\alpha}_{s,r},\{\textbf{c}_{s,r}[m]\}_{\!m=1}^{\!\tau_s}} \prod_{m=1}^{\tau_s}p\left(y_{s,r} [m] |\boldsymbol{\alpha}_{s,r},\textbf{c}_{s,r}[m],\mathcal{H}_{s,r,1}\right) p\left(\boldsymbol{\alpha}_{s,r}\right)p\left(\left\{\textbf{c}_{s,r}[m]\right\}_{m=1}^{\tau_s}|\mathcal{H}_{s,r,1}\!\right)}{ \prod_{m=1}^{\tau_s}p\left(y_{s,r} [m]|\mathcal{H}_{s,r,0}\right)}\right)
\end{align}
   \hrulefill %\vspace{-6mm}
\end{figure*}
The local test statistics are obtained following the iterative approach in \cite[Algorithm 1]{Zou2024Distributed}. For detailed derivations, we refer to \cite{Zou2024Distributed} to save space.

%\subsection{Correlation-Unaware Detector with Fully-Informed System (FIS)}
%Assuming that the correlation matrix of the RCS variables is not available and correlations among different samples are disregarded, we employ a correlation-unaware detector based on the Neyman-Pearson detector. The local test statistics is given as
%\begin{align}
 %   T_{l,s}^{\rm NP}&= \ln\Bigg( \frac{\prod_{m=1}^{\tau_s}p(y_{r,s}[m]|\mathcal{H}_{s,1})}{\prod_{m=1}^{\tau_s}p(y_{r,s}[m]|\mathcal{H}_{s,0})}\Bigg)\nonumber\\
%    &= \sum_{m=1}^{\tau_s} y_{r,s}^{*}[m] a_{r,s}^{-1}[m] y_{r,s}[m],
%\end{align}
%where $a_{r,s}[m]=\operatorname{tr}\left(\textbf{G}_{r,s}^H\textbf{v}_{r,s} \textbf{v}_{r,s}^H \textbf{G}_{r,s}\right)
%    + \sigma_n^2$. 

%The constant terms in the detector are removed since they can be handled by the threshold.

%{\color{blue}
%\subsection{Benchmark: Distributed Sensing GLRT Detector with FIS }
%\begin{align}
%    T_{l,s}^{\rm GLRT} [m]&= \sum_{m=1}^{\tau_s} \Vert \textbf{U}_{l,s}^H[m] \textbf{y}_{r,s}[m] \Vert^2,
%\end{align}
%where $\textbf{U}_{l,s}$ is the left singular eigenvectors of $\textbf{G}_{r,s}[m]$.
%\\
%OR
%\begin{align}
%    T_{l,s}^{\rm GLRT} [m]&= \sum_{m=1}^{\tau_s} \Vert \textbf{u}_{l,s}^H[m] y_{r,s}[m] \Vert^2,
%\end{align}
%where $\textbf{u}_{l,s}$ is the left singular eigenvectors of $\textbf{v}_{r,s}^H\,\textbf{G}_{r,s}[m]$.
%\\
%OR
%\begin{align}
%    T_{l,s}^{\rm GLRT} [m]&= \sum_{m=1}^{\tau_s} \Vert (\textbf{v}_{r,s}^H\,\textbf{G}_{r,s}[m])^H y_{r,s}[m] \Vert^2,
%\end{align}
%}

%%%%%%%%%%%%%%%%%%%%%%%%%%%%%%%%%%%%%%%%%%%%%%%%%%
\section{Power Allocation}
\label{sec:power}
The accuracy of local test statistics at RX-APs is influenced by sensing SINR. Lower sensing SINR reduces test accuracy and detection probability, while increasing sensing SINR raises interference for the UEs, degrading communication performance. To balance both, we maximize the minimum sensing SINR ($\gamma_s$) and communication SINR ($\gamma_c$), subject to per-AP power constraints. This leads to a multi-objective optimization problem with weights $\omega_0 ,\omega_1 > 0$, as 
\begin{subequations}
\begin{align}
& \underset{
  \boldsymbol{\rho}\geq \textbf{0}}{\textrm{maximize}} \quad  
  \omega_0\,\gamma_s+\omega_1 \gamma_c\label{obj_funcP2}\\
  & \textrm{subject to} \nonumber\\
   & \hspace{15mm}
    \mathsf{SINR}_k^{\text{comm}}
    \geq \gamma_c,\quad \forall k\label{const:comm_sinr}\\
    &\hspace{15mm}\mathsf{SINR}_{s,r}^{\rm{sens}}\geq \gamma_s, \quad \forall s , \forall r \label{const:sensing_sinr}\\
    &\hspace{15mm}\sum_{k=1}^{K}p_{k,l} + \sum_{s=1}^{S} q_{s,l} \leq P_{\rm tx},\quad \forall l\in \mathcal{L}_{\rm tx}\label{const:power}\\
    & \hspace{15mm} 0\leq p_{k,l}\leq \sqrt{P_{\text{tx}}}\,\eta_{k,l}, \quad \forall k\in \mathcal{K}, \forall l\in \mathcal{L}_{\rm tx}\label{const:zero_comm_power}\\
    & \hspace{15mm} 0\leq q_{s,l}\leq \sqrt{P_{\text{tx}}}\,\zeta_{s,l}, \quad \forall s\in \mathcal{S}, \forall l\in\mathcal{L}_{\rm tx} \label{const:zero_sens_power}.
\end{align}
\end{subequations}
The optimization problem is non-convex due to \eqref{const:comm_sinr} and \eqref{const:sensing_sinr}. To convexify these constraints, we apply CCP and introduce slack variables $\xi_{k}$ and $\chi_{s,r}$ to ensure feasibility in the initial iterations. Applying CCP to \eqref{const:comm_sinr}, at the $c$th iteration, we obtain 
\begin{align}
    &\frac{2\boldsymbol{\rho}_k^T\textbf{a}_k^*\textbf{a}_k^T\boldsymbol{\rho}_k^{(c)}}{\gamma_c^{(c)}} - \frac{\left(\boldsymbol{\rho}_k^{(c)}\right)^T\textbf{a}_k^*\textbf{a}_k^T\boldsymbol{\rho}_k^{(c)}}{\left(\gamma_c^{(c)}\right)^2} \gamma_c+\xi_k\nonumber\\
    &\geq \sum_{j\in\mathcal{K}}\boldsymbol{\rho}_j^T \textbf{B}_{kj}\boldsymbol{\rho}_j+ \sum_{s\in \mathcal{S}}\boldsymbol{q}_s^T \textbf{C}_{ks}\boldsymbol{q}_s + \sigma_n^2,\label{const:comm_CCP}
\end{align}
and applying CCP to \eqref{const:sensing_sinr}, we have 
\begin{align}\label{const:sensing_CCP}\frac{2{\boldsymbol{\rho}}^T \!\boldsymbol{\mathsf{A}}_{s,r}{\boldsymbol{\rho}}^{(c)}}{\gamma_c^{(c)}}\!- \!\frac{\left(\boldsymbol{\rho}^{(c)}\right)^T\!\boldsymbol{\mathsf{A}}_{s,r}\boldsymbol{\rho}^{(c)}}{\left(\gamma_s^{(c)}\right)^2} \gamma_s\!+ \!\chi_{s,r}\!\geq\! {\boldsymbol{\rho}}^T \boldsymbol{\mathsf{B}}_{s,r}{\boldsymbol{\rho}}\!+\!\tau_s\sigma_n^2.
\end{align}

The constraint~\eqref{const:power} represents the per-AP power constraint. The constraints \eqref{const:zero_comm_power} and \eqref{const:zero_sens_power} are to guarantee that the communication and sensing power coefficients for UE $k$ and SSA $s$ are zero if they are not assigned to TX-AP $l$. To do this, we define $\eta_{k,l}$ in \eqref{const:zero_comm_power}, where it is one if $k\in \mathcal{U}_l$ and zero otherwise. Similarly in \eqref{const:zero_sens_power}, $\zeta_{s,l}=1$ if $s\in \mathcal{S}_l$. The optimization problem is rewritten as 
\begin{subequations}
\begin{align}
\textbf{P:} \quad& \underset{
  \boldsymbol{\rho}\geq \textbf{0}, \chi_{s,l}, \xi_{k}\geq 0. }{\textrm{minimize}} \quad  
  -\omega_0\gamma_s-\omega_1 \gamma_c+ \lambda \sum_{ s, r}\chi_{s,r} +\lambda\sum_{k}\xi_k\label{obj_funcP2}\\
  & \textrm{subject to} \quad\eqref{const:comm_CCP}, \eqref{const:sensing_CCP}, \eqref{const:power},\eqref{const:zero_comm_power},\eqref{const:zero_sens_power}\nonumber.
\end{align}
\end{subequations}
The steps of the power allocation algorithm is outlined in Algorithm~\ref{alg:CCP_Power_Allocation}. \footnote{The CCP algorithm is guaranteed to converge to a stationary point of the problem under mild conditions, as established in \cite{lanckriet2009convergence}.}

\begin{algorithm}[h]
    \caption{\textbf{CCP-Based Multi-Objective Power Allocation}}
    \label{alg:CCP_Power_Allocation}
    \begin{algorithmic}[1]
        \STATE \textbf{Initialize:} Set initial power $\boldsymbol{\rho}^{(0)}$, sensing SINR $\gamma_s^{(0)}$, communication SINR $\gamma_c^{(0)}$, auxiliary variables $\xi_k, \chi_{s,r}$, tolerances $\epsilon_1>0$, $\epsilon_2>0$, $\epsilon_3>0$, and $\epsilon_4>0$, iteration counter $c \gets 0$, maximum iteration $c_\textrm{max}$, and convergence flag $converged \gets false$.
        \WHILE{not converged \AND  \, $c \leq c_{\max}$}
            \STATE Solve problem $\mathbf{P}$ using $\boldsymbol{\rho}^{(c)}$, $\gamma_s^{(c)}$, and $\gamma_c^{(c)}$.
            \STATE Update $\boldsymbol{\rho}^{(c+1)}$, $\gamma_s^{(c+1)}$, $\gamma_c^{(c+1)}$, $\xi_k$, $\chi_{s,r}$.
            \STATE If infeasible, reinitialize and retry.
          \STATE $c \gets c + 1$
            \IF{$|\gamma_s^{(c)} - \gamma_s^{(c-1)}| \!\leq \!\epsilon_1$ \algorithmicand \, $|\gamma_c^{(c)} \!- \!\gamma_c^{(c-1)}|\!\leq\!\epsilon_2$\,\algorithmicand\,  $\|\boldsymbol{\rho}^{(c)} \!-\! \boldsymbol{\rho}^{(c-1)} \|\leq \epsilon_3$\,\algorithmicand\, $\sum_{k}\xi_k + \sum_{s,r}\chi_{s,r}\leq \epsilon_4$}
                \STATE $converged \gets true$
            \ENDIF
        \ENDWHILE
        \STATE \textbf{Return:} Optimized power allocation $\boldsymbol{\rho}$, $\gamma_s$, $\gamma_c$.
    \end{algorithmic}
\end{algorithm}
\section{Numerical Results}
\label{sec:results}
In this section, we evaluate the sensing and communication performance based on the minimum detection probability and minimum SINR per UE. We consider a $500 \times 500\,\text{m}^2$ area with $L\!=\!25$ APs deployed in a grid, each equipped with $M\!=\!4$ antennas, and $K\!=\!8$ randomly placed UEs. The SSAs are $S=4$, located at coordinates $(125,125)$, $(125,375)$, $(375,125)$, and $(375,375)$. Each AP has a maximum transmit power of $1$\,W, while the uplink pilot transmission power is $0.2$\,W. Unless stated otherwise, all targets have the same RCS variance, $\sigma_{\rm rcs}^2=-5$\,dBsm. The false alarm probability is set to $0.03$, following \cite[Table 6.2-1]{3gpp_ts_22_137} for 3GPP object detection and tracking scenarios with sensing service category $4$, and the sensing symbol duration is $\tau_s=20$. The large-scale fading coefficients, probability of LOS, and the Rician factors are simulated based on the 3GPP Urban Microcell model, as defined in \cite[Table B.1.2.1-1, Table B.1.2.1-2, Table B.1.2.2.1-4]{3gpp2010further}. In Algorithm~\ref{algo:AP_selection}, $\beta_{\rm th}/\sigma_n^2= 80000\approx 49$
dB and in Algorithm~\ref{alg:CCP_Power_Allocation}, the solution accuracy parameters are set as $\epsilon_1=\epsilon_2=10^{-3}$, $\epsilon_3 =0.1$, and $\epsilon_4 =10^{-6}$, with a maximum iteration limit of $150$. The optimization weights $\omega_0$ and $\omega_1$ are $1$, unless otherwise stated. We define the minimum detection probability as the average of the minimum detection probabilities across all SSAs.
  \begin{figure}[t]
    \centering
     %\vspace{-1mm}
   \includegraphics[width=0.7\linewidth]{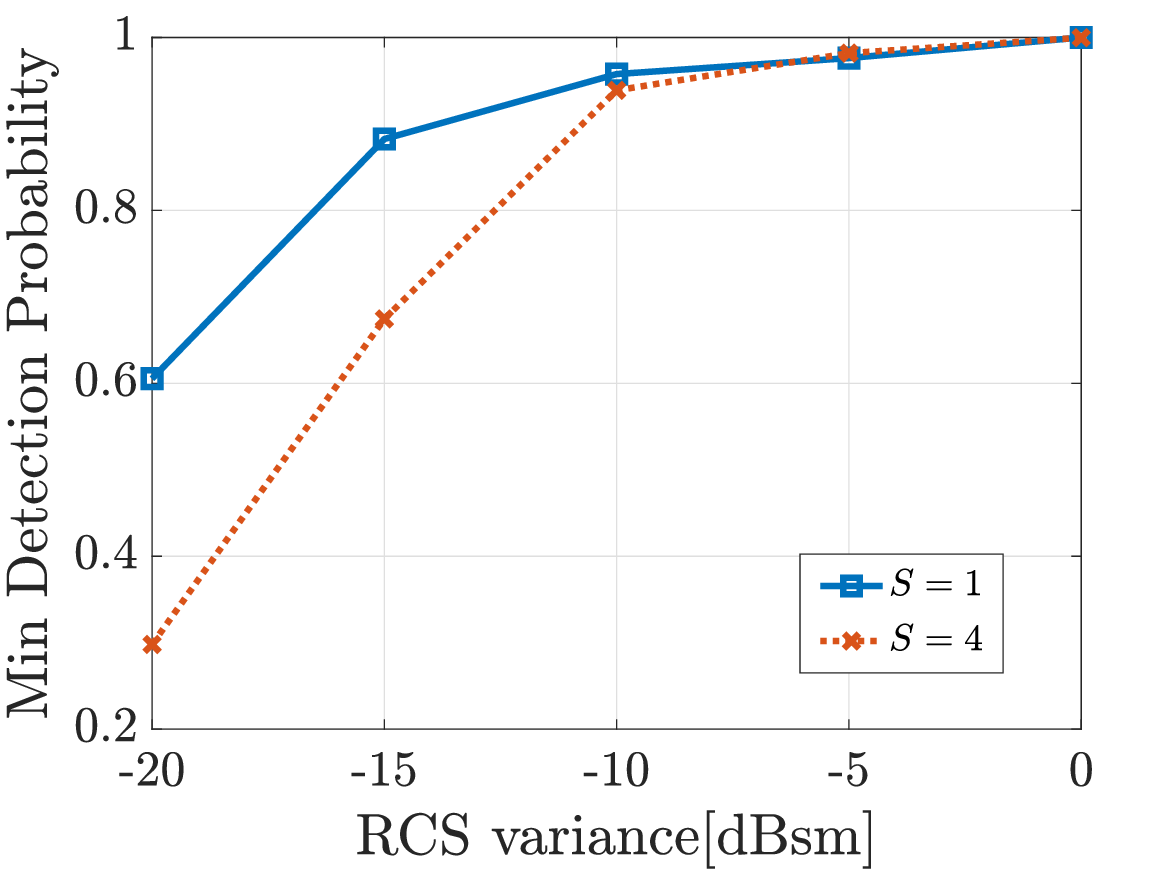} 
\caption{ Minimum detection probability vs. RCS variance for $R=T=1$ and FIS.}\label{fig:RCS} %\vspace{-6mm}
\end{figure}

Fig.~\ref{fig:RCS} depicts the minimum detection probability versus RCS variance for single-target ($S=1$) and multi-target ($S=4$) scenarios with $R=T=1$ in FIS. When the RCS variance exceeds $-10$\,dBsm, the probability remains above 0.9 for both cases. However, as RCS decreases, the gap widens: for $S=1$, it stays above 0.8, whereas for $S=4$, it drops to around 0.67 when $\sigma_{\rm rcs}^2=-15$\,dBsm.

\begin{figure*}[h!]
    \begin{subfigure}{0.33\textwidth}
        \centering
    \includegraphics[width=1\textwidth]{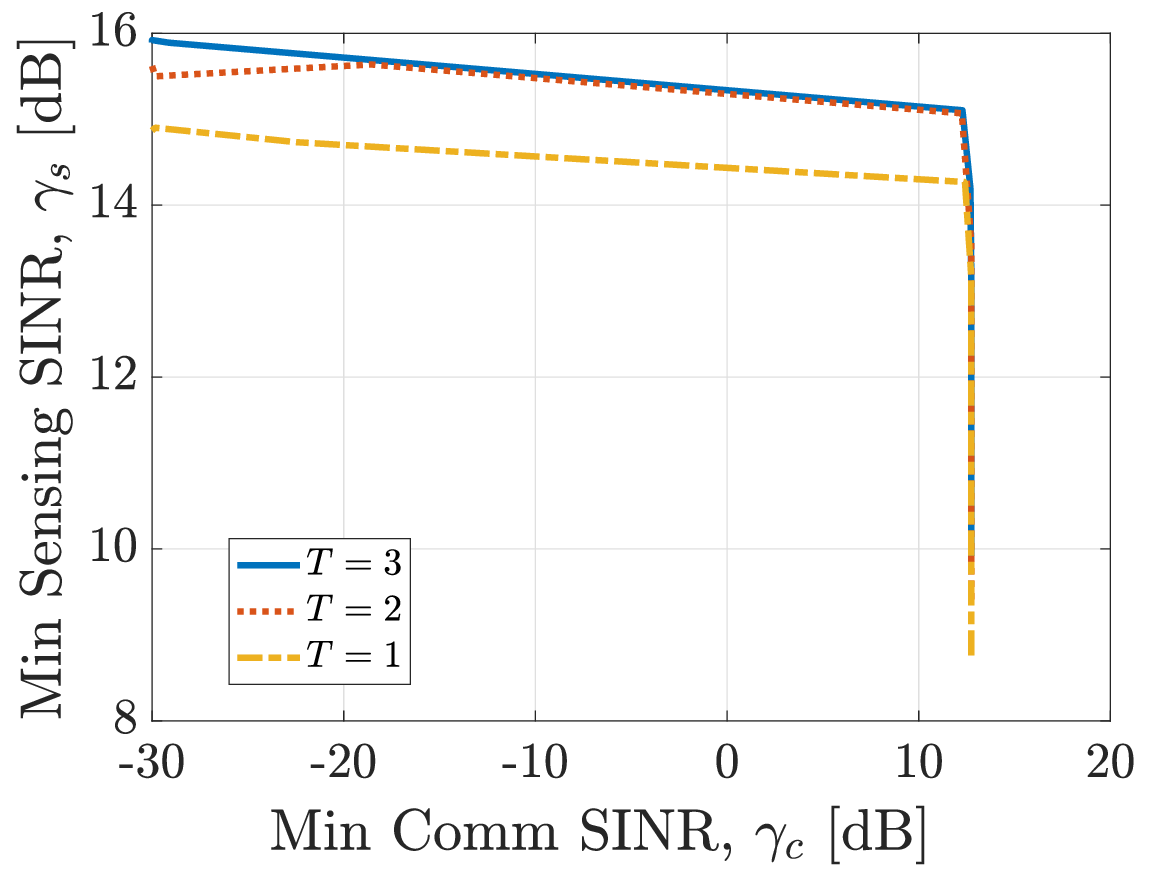} 
   \caption{} \label{fig: gammaCS}
        \end{subfigure}
        %\hfill
          \begin{subfigure}{0.33\textwidth}
    \centering
   \includegraphics[width=1\textwidth]{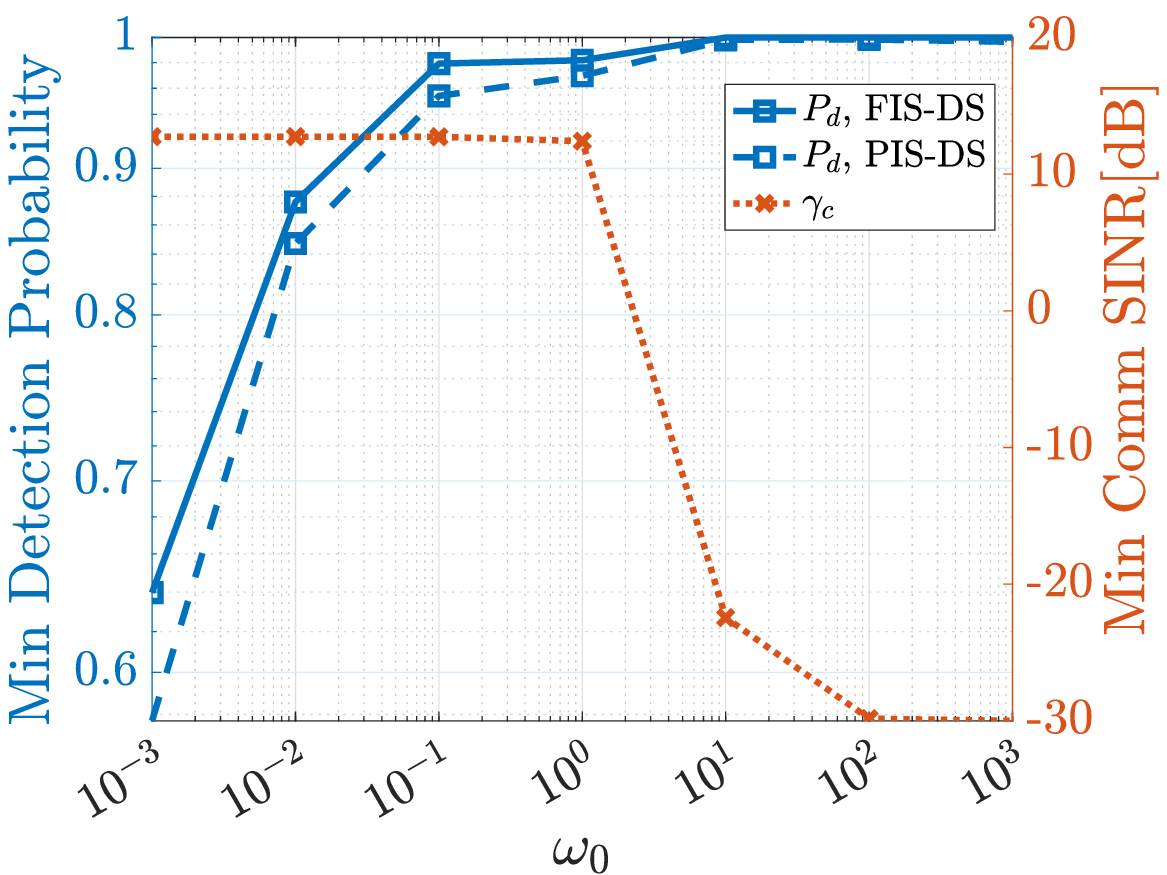} 
   \caption{} \label{fig: Pd_w0}
   \end{subfigure}
    \begin{subfigure}{0.33\textwidth}
    \centering
   \includegraphics[width=1\textwidth]{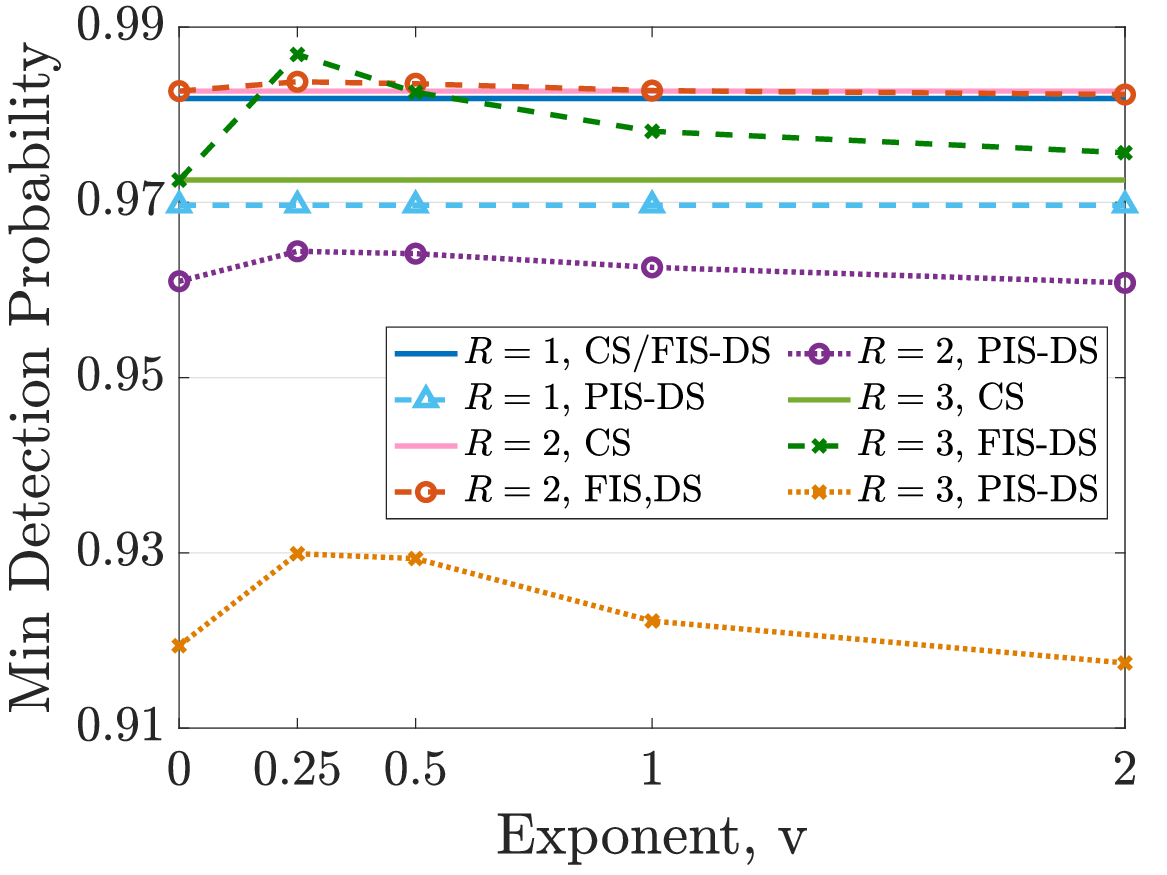} 
   \caption{} \label{fig: exponent}
   \end{subfigure}
 \caption{(a) Minimum sensing SINR vs. minimum communication SINR with $R=1$, (b) minimum detection probability with FIS and PIS and min communication SINR vs. optimization weight $\omega_0$ for $T=1$, (c) minimum detection probability vs. exponent $v$ for $S=4$ and $\omega_0=1$.} \label{fig:2}%\vspace{-6mm}
    \end{figure*}

%\begin{figure}[t]

%\centerline{\includegraphics[trim={0mm 0mm 0mm 2mm},clip,width=0.85\linewidth]{figures/Pd_T.jpg}}
%\caption{Minimum detection probability vs. number of TX-APs per sensing service area with $R=1$, $S=4$ and $\sigma_{\rm rcs} =-5$\,dBsm.}\label{fig:Tx} 
%\vspace{-4mm}
%\end{figure}

%\begin{figure}[tbp]

%\centerline{\includegraphics[trim={0mm 0mm 0mm 2mm},clip,width=0.85\linewidth]{figures/Pd_R.jpg}}
%\caption{Minimum detection probability vs.  number of RX-APs with $T=1$, and $\sigma_{\rm rcs} =-5$\,dBsm.}\label{fig:Rx} 
%\vspace{-4mm}
%\end{figure}

Fig.~\ref{fig: gammaCS} illustrates the trade-off between the minimum sensing SINR per SSA, $\gamma_s$, and the minimum communication SINR per UE, $\gamma_c$, as $\omega_0$ varies from $10^{-3}$ to $10^{3}$, with $\omega_1=1$ and $T=1,2,3$. Initially, as $\omega_0$ decreases, $\gamma_s$ gradually decreases while $\gamma_c$ improves, indicating a shift in resource allocation toward communication. However, for $\omega_0<1$, further reduction in $\gamma_s$ does not lead to significant gains in $\gamma_c$ due to communication system constraints. Additionally, increasing the number of TX-APs per SSA ($T$) enhances $\gamma_s$, as seen from the higher sensing SINR values for $T=2$ compared to $T=1$. However, adding one more TX-AP, i.e., $T=3$, does not significantly improve the sensing performance. This could be due to interference to communication UEs and power limits.

Fig.~\ref{fig: Pd_w0} shows the minimum detection probability for FIS and PIS as a function of optimization weight $\omega_0$. Both schemes achieve a detection probability of 1 for $\omega_0 \geq 10$, but at the cost of very low minimum communication SINR. However, the system can maintain a detection probability of 0.98 while ensuring a minimum UE SINR above 10\,dB when $\omega_0=1$. Moreover, comparing Fig.~\ref{fig: gammaCS} and Fig.~\ref{fig: Pd_w0}, we conclude that a minimum sensing SINR of approximately 14\,dB is required for a minimum detection probability of $0.98$.

Fig.~\ref{fig: exponent} illustrates the impact of exponent $v$ in \eqref{eq:weights} and the number of RX-APs per SSA on the minimum detection probability for $R=1,2,3$. We compare FIS and PIS for distributed sensing (DS) against centralized sensing (CS) as a benchmark \cite{behdad2022power}. Additionally, we evaluate the proposed weighted distributed approach against the unweighted case ($\!v\!=\!0$).
Weighted local measurements improve detection performance over unweighted distributed sensing. Since the first selected RX-AP typically has a stronger channel, adding more RX-APs may degrade local test statistics due to weaker channels, especially in PIS scenarios. However, FIS-DS with $R=2$ achieves slightly higher detection probability than CS/FIS-DS with $R=1$. The results show that $v=0.25$ maximizes sensing performance for $R=2$. Interestingly, FIS-DS with $R=3$ and $v=0.25$ outperforms its counterparts with $R=1$ and $R=2$, highlighting the benefits of the proposed channel-aware sensing approach and the role of exponent $v$.
Table~\ref{tab:R_gammaC} presents the minimum communication SINR for $R=1,2,3$. With a limited number of APs, increasing $R$ degrades communication performance, reducing $\gamma_c$ by approximately 10 dB as $R$ increases from 1 to 3.
\begin{table}[t]
    \centering
  %  \vspace{3mm}
\caption{Minimum communication SINR vs. $R$ }
\label{tab:R_gammaC}
    \begin{tabular}{|c|c|l|l|} \hline 
         $R$&  1& 2&3\\ \hline 
 $\gamma_c$[dB]&12.4119& 5.0597&1.2618\\\hline
 \end{tabular}
    
    %\vspace{-5mm}
\end{table}

\section{Conclusion}
\label{sec:conclusion}
In this paper, we explore multi-target detection in an ISAC-enabled CF-mMIMO system with C-RAN architecture, using distributed sensing and communication. A heuristic AP selection method is proposed for sensing, where the nearest RX- and TX-AP pairs are assigned to each service area. We also introduce a channel-aware distributed sensing scheme, weighting local test statistics based on SIR values.
Our findings show that the weighted sensing scheme outperforms non-weighted methods, improving the minimum detection probability. While increasing the number of RX-APs degrades sensing performance due to weaker channels and higher errors, this can be mitigated by selecting an appropriate exponent value. Moreover, assigning more RX-APs to the sensing areas results in communication performance degradation, up to $10$\,dB communication SINR loss, due to limited communication resources.
%\vspace{-2mm}
\bibliographystyle{IEEEtran}
\bibliography{IEEEabrv.bib,refs.bib}

\begin{thebibliography}{10}
\providecommand{\url}[1]{#1}
\csname url@samestyle\endcsname
\providecommand{\newblock}{\relax}
\providecommand{\bibinfo}[2]{#2}
\providecommand{\BIBentrySTDinterwordspacing}{\spaceskip=0pt\relax}
\providecommand{\BIBentryALTinterwordstretchfactor}{4}
\providecommand{\BIBentryALTinterwordspacing}{\spaceskip=\fontdimen2\font plus
\BIBentryALTinterwordstretchfactor\fontdimen3\font minus
  \fontdimen4\font\relax}
\providecommand{\BIBforeignlanguage}[2]{{%
\expandafter\ifx\csname l@#1\endcsname\relax
\typeout{** WARNING: IEEEtran.bst: No hyphenation pattern has been}%
\typeout{** loaded for the language `#1'. Using the pattern for}%
\typeout{** the default language instead.}%
\else
\language=\csname l@#1\endcsname
\fi
#2}}
\providecommand{\BIBdecl}{\relax}
\BIBdecl

\bibitem{liu2022integrated}
F.~Liu, Y.~Cui, C.~Masouros, J.~Xu, T.~X. Han, Y.~C. Eldar, and S.~Buzzi,
  ``Integrated sensing and communications: Toward dual-functional wireless
  networks for {6G} and beyond,'' \emph{IEEE Journal on Selected Areas in
  Communications}, vol.~40, no.~6, pp. 1728--1767, 2022.

\bibitem{demir2024cell}
{\"O}.~T. Demir, M.~Masoudi, E.~Bj{\"o}rnson, and C.~Cavdar, ``Cell-free
  massive {MIMO} in {O-RAN}: Energy-aware joint orchestration of cloud,
  fronthaul, and radio resources,'' \emph{IEEE Journal on Selected Areas in
  Communications}, vol.~42, no.~2, pp. 356--372, 2024.

\bibitem{8000355}
S.~Buzzi and C.~D’Andrea, ``Cell-free massive {MIMO}: User-centric
  approach,'' \emph{IEEE Wireless Communications Letters}, vol.~6, no.~6, pp.
  706--709, 2017.

\bibitem{cell-free-book}
{\"{O}}.~T. Demir, E.~Bj\"{o}rnson, and L.~Sanguinetti, ``Foundations of
  user-centric cell-free massive {MIMO},'' \emph{Found. Trends Signal
  Process.}, vol.~14, no. 3-4, pp. 162--472, 2021.

\bibitem{buzzi2024scalability}
S.~Buzzi, C.~D’Andrea, and S.~Liesegang, ``Scalability and implementation
  aspects of cell-free massive {MIMO} for {ISAC},'' in \emph{2024 19th
  International Symposium on Wireless Communication Systems (ISWCS)}, 2024, pp.
  1--6.

\bibitem{zhang2023multi-target}
X.~Zhang, H.~Zhang, R.~Deng, L.~Liu, and B.~Di, ``Multi-target detection for
  reconfigurable holographic surfaces enabled radar,'' in \emph{GLOBECOM 2023 -
  2023 IEEE Global Communications Conference}, 2023, pp. 6634--6639.

\bibitem{elfiatoure2024multiple}
M.~Elfiatoure, M.~Mohammadi, H.~Q. Ngo, and M.~Matthaiou, ``Multiple-target
  detection in cell-free massive {MIMO}-assisted {ISAC},'' \emph{arXiv preprint
  arXiv:2404.17263}, 2024.

\bibitem{Zou2024Distributed}
Q.~Zou, Z.~Behdad, {\"O}.~T. Demir, and C.~Cavdar, ``Distributed versus
  centralized sensing in cell-free massive {MIMO},'' \emph{IEEE Wireless
  Communications Letters}, vol.~13, no.~12, pp. 3345--3349, 2024.

\bibitem{wang2020uplink}
Z.~Wang, J.~Zhang, E.~Bj{\"o}rnson, and B.~Ai, ``Uplink performance of
  cell-free massive {MIMO} over spatially correlated rician fading channels,''
  \emph{IEEE Communications Letters}, vol.~25, no.~4, pp. 1348--1352, 2020.

\bibitem{lanckriet2009convergence}
G.~Lanckriet and B.~K. Sriperumbudur, ``On the convergence of the
  concave-convex procedure,'' \emph{Adv. in {N}eur. {I}nf. {P}roc. {S}ys.},
  vol.~22, 2009.

\bibitem{3gpp_ts_22_137}
{3rd Generation Partnership Project (3GPP)}, ``{Service requirements for
  Integrated Sensing and Communication; Stage 1 (Release 19)},'' 3GPP,
  Technical Specification TS 22.137 V19.1.0, March 2024, available online:
  \url{https://www.3gpp.org}.

\bibitem{3gpp2010further}
3GPP, ``Further advancements for {E-UTRA} physical layer aspects (release 9),''
  \emph{TS 36.814}, 2017.

\bibitem{behdad2022power}
Z.~Behdad, {\"O}.~T. Demir, K.~W. Sung, E.~Bj{\"o}rnson, and C.~Cavdar, ``Power
  allocation for joint communication and sensing in cell-free massive {MIMO},''
  in \emph{GLOBECOM 2022-2022 IEEE Global Communications Conference}.\hskip 1em
  plus 0.5em minus 0.4em\relax IEEE, 2022, pp. 4081--4086.

\end{thebibliography}
\end{document}